\documentstyle[preprint,prl,aps]{revtex}
\draft
\preprint{IMSc-97/03/08}

\def\a{\alpha}
\def\b{\beta}

\def\e{\epsilon}
                    
\def\vf{\varphi}

\def\l{\lambda}                 
\def\m{\mu}
\def\n{\nu}
                  
\def\p{\pi}                     
\def\q{\theta}                  
\def\r{\rho}
\def\s{\sigma}

\def\dsp{\displaystyle}
\def\barr{\begin{array}}
\def\earr{\end{array}}

\def\ms2{m_{B^*}^{2}}

\def\ba{\begin{array}}
\def\ea{\end{array}}
\def\by{\begin{eqnarray}}
\def\ey{\end{eqnarray}}

\def\nn{\nonumber}

\def\bkpiee{$B\to K^* \ell^+\ell^-\to K \pi \ell^+\ell^-$ }

\def\bib{\bibitem}
\def\be{\begin{equation}}
\def\ee{\end{equation}}
\def\beqar{\begin{eqnarray}}
\def\eeqar{\end{eqnarray}}
\def\barr{\begin{array}}
\def\earr{\end{array}}

\def\and{\qquad {\rm and } \qquad}

\def\npb#1{Nucl. Phys. {\bf B#1}}

\begin{document}
\title{$C\!P$~ violating anomalous trilinear gauge couplings from
$B\to K^* \ell^+ \ell^-$}
\author{Nita Sinha and Rahul Sinha
\ifpreprintsty
\footnote{e-mail: nita,sinha@imsc.ernet.in}
\else
\cite{author}
\fi
}
\address{ Institute of Mathematical Sciences, Taramani, Chennai 600113, 
India.}
\date{22 July 1997}
\maketitle
\begin{abstract}            
  We study the contributions of the $C\!P$~ violating anomalous
  $WW\gamma$ interactions to $b\rightarrow s \ell^+\ell^-$. We obtain
  cutoff independent results on $\tilde\kappa$ and $\tilde\lambda$, by
  constructing an asymmetry for the process $B\rightarrow
  K^{*}\ell^+\ell^{-}$, where the $B$ and $\bar B$ events are added.
  We show that a sample of $10^4$ $B\rightarrow K^{*}e^+e^{-}$ events
  can yield a bound, $|\tilde\kappa| < 0.42$ at 90\% C.L., which is
  much tighter than the recent constraint from D0.
\end{abstract}

Two of the aspects of the Standard Model(SM) which remain to be
further explored are the gauge nature of the W, Z bosons and $C\!P$~
violation. The vector boson self interactions are uniquely fixed by
the $SU(2)_L\times U(1)$ gauge structure of the SM. However, new
physics as well as higher order corrections will modify these self
interactions. While tests of trilinear gauge couplings are now being
done at present colliders, only weak bounds on them exist\cite{bounds
CPV,bounds CPC}. $C\!P$~ violation on the other hand is only
parameterized in the SM via the CKM matrix, and has been seen only in
the neutral K mesons. Much effort is now being put into studying
$C\!P$~ violation in the $B$ mesons. Large number of $B$ mesons are
expected to be produced in the near future, at the new asymmetric
colliders and upgraded Tevatron, which would enable study of its rare
decay modes. Rare decays of $B$ mesons, occur in the SM only through
loops and are thus expected to be very sensitive to new physics. In
particular the mode $b \rightarrow s \ell^+\ell^-$ has been
extensively studied\cite{Hewett,Wyler,Burdman2} for possible
contributions from new physics. New physics that affects the gauge
nature of the W, Z bosons, can be parameterized in terms of $C\!P$~
conserving as well as $C\!P$~ violating anomalous trilinear gauge
boson couplings. The $C\!P$~ conserving couplings have been considered
in the rare modes $b \rightarrow s\gamma$, $b \rightarrow s
\ell^+\ell^-$\cite{He,Drew,Numata}. Since, the origin of $C\!P$~
violation is not understood, any potential source of $C\!P$~ violation
should be pursued. Here we focus attention on the $C\!P$~ violating
anomalous couplings, by studying the mode $B \rightarrow K^{*}
\ell^+\ell^-$.

        The couplings of W bosons to the photon and Z can be described
by an effective Lagrangian, the most general form of which may be
written down with the minimum requirements-- that of Lorentz
invariance, global $SU(2)$ and local $U(1)$ symmetry
\cite{peccei} as:
\begin{equation} 
\begin{array}{rcl}
\displaystyle
     {\cal L}_{\it eff}^{V}= &-& g_{V}
             \left[i g_1^V
               \left( W^\dagger_{\alpha \beta} W^\alpha
                      - W^{\dagger\alpha} W_{\alpha \beta}
                \right) V^\beta
             +
              i \kappa_V  W^\dagger_{\alpha} W_\beta
                               V^{\alpha\beta}\right.\\[3ex]
            &+&\displaystyle \left .            
                  i \frac{\lambda_V}{M_W^2}
                 W^\dagger_{\alpha \beta} {W^\beta}_\sigma
                 V^{\sigma\alpha} 
            +
              i \tilde{\kappa}_V  W^\dagger_{\alpha} W_\beta
                               \tilde{V}^{\alpha\beta}
            +  i\frac{\tilde{\lambda}_V}{M_W^2}
                 W^\dagger_{\alpha \beta} {W^\beta}_\sigma
                 \tilde{V}^{\sigma\alpha} \right.\displaystyle\\[3ex]
            &+& \left. g_4^V  W^\dagger_{\alpha} W_\beta
                 (\partial^\alpha V^\beta+\partial^\beta V^\alpha) 
            + g_5^V \epsilon^{\alpha\beta\mu\nu} W^\dagger_\alpha
\stackrel{\leftrightarrow}{\partial_\mu} W_\beta V_\nu 
 \right].
      \label{lagrangian}
\end{array}
\end{equation} 
In the above equation $V$ represents the neutral gauge bosons either
the photon or the Z, $V_{\alpha\beta} = \partial_\alpha V_\beta -
\partial_\beta V_\alpha $, $W_{\alpha\beta} = \partial_\alpha W_\beta-
\partial_\beta W_\alpha $,
$\tilde{V}^{\alpha\beta}=\frac{1}{2}\epsilon^{\alpha\beta\rho\sigma}
V_{\rho\sigma}$ and $g_{V}$ is the $WWV$ coupling strength in the SM
with $g_{\gamma}=e$ and $g_{Z}=ec/s$, where $c^2 \equiv 1 -s^2 \equiv
M_W^{2}/M_Z^2$. In the SM, the couplings $g_1^V$, $\kappa_V$ are unity
and all others are zero. New physics may result in a modification of
these couplings. The deviation from the SM, the so called anomalous
couplings, need to be constrained experimentally.  We shall here
concentrate on the $C\!P$~ violating couplings $\tilde\kappa$ and
$\tilde\lambda$. While many theoretical and experimental studies of
anomalous triple gauge boson couplings have been done, few bounds
exist on the anomalous $C\!P$~ violating couplings. Direct bounds have
recently been obtained\cite{bounds CPV} on $\tilde\kappa$ and
$\tilde\lambda$ from $p\bar p\to W \gamma X$ at the tevatron.
Stringent indirect constraints\cite{Marciano} on these couplings come
from electric dipole moment (EDM) of the neutron and electron.
However, these constraints assume naturalness and are cutoff
dependent. Hence, bounds from $C\!P$~ violating asymmetries that we
pursue here, would be complementary to the EDM bounds. Possible
contributions of the couplings $\tilde\kappa$ and $\tilde\lambda$ to
processes at colliders have also been examined for LEPII\cite{peccei},
upgraded tevatron\cite{Valencia} as well as future linear
colliders\cite{Colliders}. In the chiral Lagrangian approach the
dimension 6 operator with coefficient, $\tilde\lambda$, is ignored.
For `$b$' penguin modes such as $b\rightarrow s\gamma$ or
$b\rightarrow s \ell^+\ell^-$, unitarity of the CKM, results in cutoff
independence for contributions from $\tilde\kappa$ and
$\tilde\lambda$. Since both $\tilde\kappa$ and $\tilde\lambda$
contributions are finite, requiring no cutoff, we do not neglect
$\tilde\lambda$. However, the constraints on $\tilde\lambda$ are
expected to be weaker.

The effective short distance Hamiltonian relevant to the decay $b\to
s\ell^+\ell^-$ \cite{Buras-Munz,Grinstein,ODonnell} leads to the QCD
corrected matrix element,
\begin{equation}
{\cal M}
(b\to s \ell^+\ell^-)
=\frac{\alpha G_F}{\sqrt{2}\pi}  v_t \{
-2 i C_7 m_b \frac{q^\nu}{q^2}\;\bar{s}\sigma_{\mu\nu}b_R\; 
\bar{\ell}\gamma^\mu\ell 
+C_8\;\bar{s}\gamma_\mu b_L\;\bar{\ell}\gamma^\mu\ell 
+C_9\;\bar{s}\gamma_\mu b_L\;\bar{\ell}\gamma^\mu\gamma_5\ell 
\},
\label{heff}
\end{equation}
where only the dominant top quark contribution to the loop is retained.
$C_{j}$ ($j=7,8,9$) are the Wilson coefficients given in 
Ref.\cite{Buras-Munz,Grinstein}, $m_b$ is the mass of the $b$ quark, $q^2$  
is the invariant lepton mass squared, $b_{L,R}=(1\mp\gamma_5)/2\,b$ and 
$v_t=V^*_{ts}V_{tb}$ is the product of the CKM matrix elements.
The anomalous WWV couplings, result in a shift in the values of the
short distance coefficients $C_{j}$'s at $\mu=M_{W}$, 
\begin{equation}
        C_{j} = C_{j}^{SM}+\imath\;\tilde\kappa
        C_{j}^{\tilde\kappa}+\imath\;\tilde\lambda
        C_{j}^{\tilde\lambda},
\end{equation}
where, $C_{j}^{SM}$ are the SM coefficients and
$C_{j}^{\tilde\kappa}$, $C_{j}^{\tilde\lambda}$ are the contributions
from the anomalous couplings. These shifted values of $C_j$'s are then
evolved down to the $b$ quark scale. The anomalous $WW\gamma$ coupling
coefficients have been evaluated to be,
\begin{eqnarray}
C_{8}^{\tilde\kappa}=C_{8}^{\tilde\lambda}=
  \frac{m_b^2}{M_W^2}\,\frac{x_t}{12}\,
   \left( {{2 + 37\,{x_t} + 10\,{x_t^2} - {x_t^3}}\over
       {6\,
         {{\left( 1 - x_t \right) }^4}}} +
     {{{x_t}\,\left( 3 + 5\,{x_t} \right) \,\ln {x_t}}\over
       {
         {{\left( 1 - x_t \right) }^5}}} \right),\nonumber 
\end{eqnarray}
\begin{eqnarray} 
C_{7}^{\tilde\kappa}=
\frac{ {x_t}}{2}   \left({ { 1 }\over 
       {{\left( 1 - x_t \right) }^2}
         } +
     {{{x_t}\,\left( 3 - {x_t} \right) \,
         \ln {x_t}}\over
       {2 {{\left( 1 - x_t \right) }^3}
   }} \right), 
\end{eqnarray}
\begin{eqnarray} 
C_{7}^{\tilde\lambda}=
\frac{{x_t}}{2}\,\frac{\left( 1 - {x_t^2} + 2\,{x_t}\,\ln {x_t}
\right) } 
    {2\,{{\left( 1 - x_t \right) }^3}},\nonumber
\end{eqnarray}
where, $x_t=\displaystyle\frac{m_t^2}{M_W^2}$, $m_t$ is the mass of
the top quark.  The WWZ coefficients can be related to the
corresponding $WW\gamma$ coefficients listed above, by modifying the
couplings and propagator. $C_{9}^{\tilde\kappa,\tilde\lambda}$ also
contributes to the WWZ vertex, with a functional form in $x_t$,
identical to that of $C_{8}^{\tilde\kappa,\tilde\lambda}$. However,
the anomalous contributions from the WWZ vertex may be neglected in
comparison to that of the photon. Our results for
$C_{7}^{\tilde\kappa}$ are in agreement with those of refs.\cite{He}
and \cite{Numata} but we disagree with ref.\cite{Numata} on
$C_{7}^{\tilde\lambda}$ and on $C_{8}^{\tilde\lambda}$.
$C_{8}^{\tilde\kappa}$ has been evaluated for the first time. Note
that $C_{8}^{\tilde\kappa}$ turns out to be exactly equal to
$C_{8}^{\tilde\lambda}$, perhaps indicating that for the contact term
$C_8$, the derivatives from dimension 6 operator and the gauge part of
the propagators contributing in case of the dimension 4 operator,
behave similarly. Another interesting feature is that $C_8$ is doubly
chirally suppressed for the $C\!P$~ violating couplings
$\tilde\kappa,\tilde\lambda$.

Previous studies\cite{He,Numata} of $C\!P$~ violating anomalous gauge
boson couplings in rare B decays, have used the mode $b\to s\gamma$,
to constrain $\tilde\kappa$ and $\tilde\lambda$ separately,
disallowing any possible cancelations between them. They also used
only the decay rate of $b\to s\gamma$ and hence depended quadratically
on these couplings. In order to achieve maximum sensitivity to $C\!P$~
violating couplings and to have a clear signal of $C\!P$~ violation,
we need to estimate observables that are explicitly $C\!P$~ odd and
must therefore depend linearly on the $C\!P$~ violating anomalous
couplings. $C\!P$~ violating asymmetries have also been considered in
$b\to s\gamma$, in SM and two Higgs doublet
model\cite{Soares,Wolfenstein,chinese}. However, such asymmetries rely
on the presence of a large strong phase arising out of final state
interactions. $C\!P$~ violating asymmetries in rare modes that require
flavour tagging are going to be very difficult to detect. We therefore
prefer to use a technique that neither needs flavour nor time tagging.
Such an asymmetry has been considered in ref.\cite{Sinha}. We adopt
this technique and construct asymmetries for the mode $B \rightarrow
K^{*}\ell^{+}\ell^{-}$ ($\ell^+\ell^-$ non resonant), to obtain
simultaneous bounds on $\tilde\kappa$ and $\tilde\lambda$. The mode $B
\rightarrow K^{*}\ell^+\ell^{-}$ is theoretically clean and can
provide additional information due to its richer kinematics. We
present here only the case of charged $B$'s. For neutral $B$'s, after
time integration, the asymmetries that we construct, have exactly the
same form as that for the charged $B$'s \cite{Sinha}.

The transition matrix element for the exclusive process $B(p)\to K^*(k) 
\ell^+\ell^-\to K(k_1) \pi (k_2)\ell^+(q_1)\ell^-(q_2)$ can be written 
for each of the operators in Eq.(\ref{heff}) as,
\begin{eqnarray}
\langle K\pi|\bar{s}i\sigma_{\mu\nu}(1\pm\gamma_5)q^\nu b|B\rangle &=& 
i{\cal A} \epsilon_{\mu\nu\alpha\beta} K^\nu k^\alpha q^\beta\pm {\cal 
B}K_\mu\pm {\cal C} k_\mu \, ,\nonumber \\
\langle K\pi|\bar{s}\gamma_\mu(1\mp\gamma_5)b|B\rangle &=& i{\cal 
D}\epsilon_{\mu\nu\alpha\beta} 
K^\nu k^\alpha q^\beta \pm {\cal E} K_\mu\pm {\cal F} k_\mu \,.
\end{eqnarray}
The form-factors ${\cal A,\cdots,F}$ are unknown functions of
$q^2=(p-k)^2$ and other dot products involving momentum, $k=k_1+k_2$
and $K=k_1-k_2$ and can be related to those used in \cite{Wyler}, as
given in the Table. The variable $\sigma$, in the table, arises due to
the decay of $K^*\to K\pi$, evaluated in the zero width approximation.
They can be similarly related to those in heavy quark effective theory
(HQET)\cite{Burdman2}. The HQET form factors cannot currently be
reliably predicted over the entire dilepton mass range. The results
that we shall obtain, do depend on the numerical values of the form
factors. In future, with large number of $B$'s available, the form
factors can be determined experimentally. The current proportional to
$q_\mu$ does not contribute as it couples to light leptons. In our
notation $M_B,m_{K^*},m_K$ and $m_\pi$ are the masses of the $B$,
$K^*$, $K$ mesons and the pion respectively. The matrix element for
the process $B\to K^* \ell^+\ell^-\to K \pi \ell^+\ell^-$ can be
written as
\begin{equation}
{\cal M}( B \displaystyle\to
K\pi\ell^+\ell^-)=\frac{\alpha G_F}{\sqrt{2}\pi}
\left\{\left(i \alpha_L\,\epsilon_{\mu\nu\alpha\beta}\, K^\nu k^\alpha 
q^\beta\right.\right. 
  \displaystyle\left. \left.+ \beta_L\, K_\mu+ \rho_L\, k_\mu
\right){\bar\ell}\gamma^\mu\,L\,\ell+ L\to R
\right\}\,,
\end{equation}
where $L,R=\displaystyle\frac{(1\mp\gamma_5)}{2}$, $q=q_1+q_2$ and 
$Q=q_1-q_2$, and the coefficients $\alpha_{L,R},~\beta_{L,R}$ and 
$\rho_{L,R}$ are given by
\widetext
\begin{equation}
\begin{array}{rclcrcl}
\displaystyle\alpha_{R,L} &=&\displaystyle 
v_t\,\{\frac{(C_8\pm C_9)}{2}\,{\cal D}
       -\frac{m_b}{q^2}\,C_7\,{\cal A}\}&=&\displaystyle|{\rm
a}_{R,L}|\,exp\left(i\delta^{\alpha}_{R,L}\right)\,exp 
\left(i\phi^{\alpha}_{R,L}\right) 
\\[2ex] 
\displaystyle\beta_{R,L}& =&\displaystyle v_t\,\{\frac{(C_8\pm 
C_9)}{2}\,{\cal E}
       -\frac{m_b}{q^2}\,C_7\,{\cal B}\}&=&\displaystyle|{\rm
b}_{R,L}|\,exp\left(i\delta^{\beta}_{R,L}\right)\,
exp\left(i\phi^{\beta}_{R,L}\right) 
\\[2ex]    
\displaystyle\rho_{R,L} &=&\displaystyle v_t\,\{\frac{(C_8\pm 
C_9)}{2}\,{\cal F} 
       -\frac{m_b}{q^2}\,C_7\,{\cal C}\}&=&\displaystyle|{\rm
r}_{R,L}|\,exp\left(i\delta^{\rho}_{R,L}\right)\,
exp\left(i\phi^{\rho}_{R,L}\right) \,.
\end{array}
        \label{alpha_beta_rho}
\end{equation} 
In the above equation $\alpha$, $\beta$ and $\rho$ are recast in terms 
of $a$, $b$ and $r$ so as to identify the strong phases $\delta$ and the 
weak phases $\phi$. Using $CPT$  invariance, the matrix element for the 
decay ${\bar B}\to\bar K \bar\pi\ell^+\ell^-$ can be obtained from the 
$B\to K\pi\ell^+\ell^-$ by replacing $\alpha_{L,R}\to 
-{\bar\alpha}_{L,R}, \beta_{L,R}\to {\bar\beta}_{L,R}, 
\rho_{L,R}\to{\bar\rho}_{L,R}$ \cite{valencia2,Okubo2}, where  
\begin{equation}
\displaystyle{\bar \alpha}_{R,L}=\displaystyle|{\rm
a}_{R,L}|\,exp\left(i\delta^{\alpha}_{R,L}\right)\, 
exp\left(-i\phi^{\alpha}_{R,L}\right) 
        \label{alpha_beta_rho_bar}
\end{equation}
and similar relations hold for $\bar\beta$ and $\bar\rho$. The matrix
element mod. squared for the process \bkpiee is worked out retaining
the imaginary parts in $\a,\b$ and $\r$ to be,
\be
\barr{rl}
\left|{\cal M}(B(p)\right.&\dsp \left.\to
K(k_1)\p(k_2)\ell^+(q_1)\ell^-(q_2))\right|^2=
\frac{\a^2 G_F^2}{2\p^2}  \\
\left\{\right. &\dsp 2\,\e_{\m\n\r\s}k^\m K^\n q^\r Q^\s
   \left( K\cdot Q\,{\rm Im}(\a _{L}\,\b_{L}^* +
    \a _{R}\,\b _{R}^*)+
      {\rm Im}(\r _{R}\,\b _{R}^* -\r _{L}\,\b _{L}^*)
 \right.  \\&\dsp\left.
 -\,k\cdot Q\,{\rm Im}(\r _{L}\,\a _{L}^*+
   \r _{R}\,\a _{R}^*) \right)  +
  2\, {\rm Re}(\r _{L}\,\a _{L}^* -\r _{R}\,\a _{R}^*) \\   &\dsp
  \left( -k\cdot K  \,q^2\,k\cdot Q  +
     k\cdot q\,k\cdot Q\,K\cdot q + m_{K^*}^2\,q^2\,K\cdot Q -
     {k\cdot q}^2\,K\cdot Q \right) \,
      \\  +&\dsp
  2\,{\rm Re}(\r _{L}\,\b _{L}^*+\r _{R}\,\b _{R}^*)\,
 \left( - k\cdot K q^2 + k\cdot q\,K\cdot q -
     k\cdot Q\,K\cdot Q \right)
       \\ + &\dsp
  2\,{\rm Re}(\a _{R}\,\b _{R}^* -\a _{L}\,\b _{L}^*)
\left( K^2\,q^2\,k\cdot Q - k\cdot Q\,{K\cdot q}^2 -
     k\cdot K\,q^2\,K\cdot Q \right.
      \\  + &\dsp \left.
 k\cdot q\,K\cdot q\,K\cdot Q \right) +
  \left( \r_L^2+\r_R^2 \right)\,
\left( -m_{K^*}^2\,q^2 + k\cdot q^2 -k\cdot Q^2 \right)
      \\  + &\dsp
  \left( {\a_L}^2 + {\a_R}^2\right)\,
\left( -K^2\,q^2\,{k\cdot Q}^2   +
     {k\cdot Q}^2\,{K\cdot q}^2+
    2\,k\cdot K\,q^2\,k\cdot Q\,K\cdot Q \right.
       \\  &\dsp -\left.
     2\,k\cdot q\,k\cdot Q\,K\cdot q\,K\cdot Q -
      m_{K^*}^2\,q^2\,{K\cdot Q}^2 +
     {k\cdot q}^2\,{K\cdot Q}^2 \right)
    \\ +&\dsp \left.
  \left( {{\b_{L}}^2} +{{\b_{R}}^2} \right)
 \left( -K^2\,q^2+{K\cdot q}^2 -
     {K\cdot Q}^2 \right) \right\}
\earr
      \label{Matmodsq}
\ee
where 
\by
k\cdot K &=& m_K^2-m_\p^2\,,\, q\cdot Q\, =\, 0  \nn \\
k\cdot q &=& \frac{1}{2}(M_B^2-m_{K^*}^2-q^2) \nn \\
k\cdot Q &=&  X M_B \cos\q_e \nn \\
K\cdot q &=&  \l_K\, X M_B \cos\q_{K^*} +
   \frac{m_K^2-m_\pi^2}{m_{K^*}^2}\, k\cdot q \nn \\
K\cdot Q &=& k\cdot Q \frac{m_K^2-m_\pi^2}{m_{K^*}^2}+
    \l_K\,( k\cdot q \cos\q_l \cos\q_K -\sqrt{q^2 m_{K^*}^2}
    \sin\q_l \sin\q_K \cos\vf ) .\nn
\ey
Here $X$ is the three momentum of the $\ell^+\ell^-$ or $K\pi$
invariant system in the $B$ meson rest frame and is given by
\[
X=\frac{(k\cdot q^2-q^2 m_{K^*}^2)^\frac{1}{2}}{M_B} 
\]
$\l_K$ is related to the $K$ three momentum in the $K^*$ rest frame and
is defined as
\[
\l_K=\left(1-\frac{(m_K+m_\p)^2}{m_{K^*}^2}\right)^\frac{1}{2}
   \left(1-\frac{(m_K-m_\p)^2}{m_{K^*}^2}\right)^\frac{1}{2} 
\]
and similarly, $\lambda_e$ is related to the lepton momentum in the
$\ell^+\ell^-$ rest frame and is given by,
\[
\lambda_e=\sqrt{1-\frac{4\,m_e^2}{q^2}}
\]
\[
\e_{\m\n\r\s} k^\m K^\n q^\r Q^\s = -X M_B \l_K \sqrt{q^2 m_{K^*}^2}
    \sin\q_l \sin\q_K \sin\vf\;,
\]
$\q_l$ ($\q_k$) is the angle between the $\ell^-$ ($K$) three-momentum
vector in the $\ell^+\ell^-$ ($K\p$) rest frame and the direction of
total $\ell^+\ell^-$ ($K^*$) three-momentum vector defined in the $B$
rest frame.  $\vf$ is the angle between the normals to the planes
defined by $\ell^+\ell^-$ and the $K\p$, in the $B$ rest frame. The
differential decay rate is then given by
\begin{equation}
d\Gamma=\displaystyle\frac{1}{2^{14} \pi^6 M_B^2}\int|{\cal M}|^2 
X\lambda_K \lambda_e\, dq^2 d\cos\theta_K d\cos\theta_l d\varphi\,,
\end{equation}
assuming a narrow width approximation for the decay $K^*\to K\pi$. 

It can easily be seen from eq.(\ref{Matmodsq}) that, the only terms
proportional to $\sin(\varphi)$ or $\sin(2 \varphi)$ are those that
depend on the imaginary parts of the products any two of
$\alpha,~\beta$ or $\rho$. For instance only the coefficient of ${\rm
Im} (\alpha_L\beta^*_L +\alpha_R\beta^*_R)$ is proportional to $\sin(2
\varphi)$. Hence we can isolate this term by considering the following
asymmetric width in terms of the differential decay rates of the $B$
meson with respect to $\varphi$,
\begin{equation}
\displaystyle \Delta \Gamma_1=\displaystyle \
 (\displaystyle\int_0^{\frac{\pi}{2}}
-\int_{\frac{\pi}{2}}^\pi
+\int_\pi^{\frac{3\pi}{2}}
-\int_{\frac{3\pi}{2}}^{2\pi})\frac{d\Gamma}{d\varphi}d\varphi\,.
\end{equation}
The imaginary part in the term under consideration can be due to
either a strong phase or a weak phase. Such $C\!P$~violating
asymmetries can be obtained not by considering the difference of
differential rates for $B$ and ${\bar B}$ , but the sum of these
rates. It follows trivially from eqn.(\ref{alpha_beta_rho} and
\ref{alpha_beta_rho_bar}) that the asymmetric width for $B$(${\bar B}$
) is,
\begin{equation}
\Delta \Gamma_1( \Delta {\bar \Gamma_1})   \propto\pm\displaystyle
\sum_{j,k} \{\displaystyle  
|a^j_L| |b^k_L| \sin(\displaystyle(\delta_L^{jk})\pm(\phi_L^{jk}))+L
\to R \},
\label {dgamma} 
\end{equation}
where $\delta_L^{jk}\equiv(\delta_L^{\alpha_j}-\delta_L^{\beta_k})$
and $\phi_L^{jk}\equiv(\phi_L^{\alpha_j}-\delta_L^{\beta_k})$. We
define the asymmetry ${\sf A}_1$ as the sum of the asymmetric widths,
normalized to the total widths, ${\sf A}_1=\displaystyle\frac{ \Delta
  \Gamma_1 + \Delta {\bar \Gamma_1}}{\Gamma + \bar \Gamma}$ and from
eq.(\ref{dgamma}), we have,
\begin{equation} 
{\sf A}_1\propto\displaystyle \sum_{j,k} \{\displaystyle |a^j_L| 
|b^k_L| \cos(\displaystyle\delta_L^{jk})\,\sin(\phi_L^{jk})+L \to R \} 
\,,\label{asymm} 
\end{equation} 
which is {\em nonzero if and only if there is $C\!P$~violation
  represented by non-zero phases
  $\phi$}\cite{Sinha,valencia2,Burdman}. $\delta$ can arise from
electromagnetic final state interactions, which are negligible and
ignored. For top quark in the penguin loop, the case of $\delta$ being
nonzero due to intermediate quark on shell, does not arise.

It is also possible to construct a different asymmetry that isolates
another combination of the imaginary terms. Such an asymmetry
\cite{kramer-palmer} considers the difference distribution of the same
hemisphere and opposite hemisphere events, and the asymmetric width in
this case can be defined by,
\begin{equation}  
\displaystyle \Delta \Gamma_2= \displaystyle 
 (\displaystyle\int_0^\pi-\int_\pi^{2\pi}) 
 d\varphi\displaystyle\int_Dd\cos\theta_l\int_Dd\cos\theta_K 
 \displaystyle\frac{d\Gamma}{d\cos\theta_l
d\cos\theta_K  d\varphi}\;,
\end{equation}
where $\displaystyle\int_D\equiv\displaystyle\int_{-1}^0-\int_0^{1}$.
Analogous to ${\sf A_1}$, we define the asymmetry ${\sf A}_2$, ${\sf
  A}_2=\displaystyle\frac{\Delta \Gamma_2+ \Delta {\bar
    \Gamma_2}}{\Gamma + {\bar \Gamma}}$.  The asymmetries ${\sf A_1}$
and ${\sf A_2}$ are evaluated to be,
\begin{equation}
{\sf A}_1=\displaystyle -2\,x\,\Delta\,\int\,dq^2 
\,{\sf C}\, ({a_0\,V-A_0\,g})\,X^2\, ,\;
{\sf A}_2 =\displaystyle\,x\,\int\,dq^2 \,{\sf 
C}\,{\sf F} \frac{1}{m_{K^*}\,\sqrt{q^2}}\,X^2\;,
\end{equation}
where, \by x&=&\displaystyle\frac{v_t^2\,\alpha^2 G_F^2\,m_b }{24 \,
  \pi^6 M_B
  (\Gamma+\bar\Gamma)(M_B+m_{K^*})}\;,\nn \\
{\sf C}&=&\displaystyle (\tilde\kappa C_7^{\tilde\kappa}+\tilde\lambda
C_7^{\tilde\lambda}) {C_8^{SM}},\;
\Delta=\displaystyle M_B^2-m_{K^*}^2, \nn \\
{\sf F}&=&\displaystyle \left\{2 X^2 M_B^2 ({a_{+}\,V-A_{+}\,g})+
  ({a_0\,V-A_0\,g})\,\Delta\,{k\cdot q}\right\}.\nn 
\ey 

Note that these asymmetries are independent of $C_9^{SM}$. Since,
$C_{8}^{\tilde\kappa,\tilde\lambda}$ terms are suppressed in
comparison with $C_{7}^{\tilde\kappa,\tilde\lambda}$, for all our
estimations, we ignore them. 

For a rate asymmetry {\sf A} to provide a $n$ standard deviation
signal of $C\!P$~ violation, we require that ${\sf A}=n/\sqrt{N}$,
where N is the total number of events in the channel.  Thus for ${\sf
A_2}$, at 1$\sigma$ (1.64 $\sigma$) level, it is possible to place the
bounds, $|\tilde\kappa |<0.34$ (0.92) for $\tilde\lambda=0$ and
$|\tilde\lambda |< 1.02$ (2.75) for $\tilde\kappa=0$, using 10,000 $(B
\rightarrow K^{*}e^{+}e^{-})$ events. This corresponds to 2x$10^9$
$B$'s, however, by adding the $(B \rightarrow K^{*}\mu^{+}\mu^{-})$
mode, $10^9$ $B$'s will be required.  Much weaker bounds would be
obtained from ${\sf A_1}$.

However, we can improve the statistical significance, by looking at
the $q^{2}$ distributions, shown in Fig. 1. We bin the asymmetry ${\sf
  A_2}$ (${\sf A_1}$ being small is ignored, and we shall drop the
subscript 2 hereafter) and use a $\chi^{2}$ fit to obtain bounds on
the parameters $\tilde\kappa$ and $\tilde\lambda$.  The asymmetry as a
function of $q^2$ is given by,
\begin{equation}  
\displaystyle A (q^2)= \frac{\displaystyle 
 (\displaystyle\int_0^\pi-\int_\pi^{2\pi}) 
 d\varphi\displaystyle\int_Dd\cos\theta_l\int_Dd\cos\theta_K 
 \;\Gamma_{sum}(q^2)}
{\displaystyle 
 (\displaystyle\int_0^\pi+\int_\pi^{2\pi}) 
 d\varphi\displaystyle\int_{-1}^1 d\cos\theta_l\int_{-1}^1 d\cos\theta_K 
 \;\Gamma_{sum}(q^2)}\;,
\end{equation}
where $\Gamma_{sum}(q^2)=\displaystyle\frac{d(\Gamma+\bar\Gamma)}{d q^2
d\cos\theta_l d\cos\theta_K d\varphi}$, $d(\Gamma+\bar\Gamma)$ is the
sum of the differential widths of $B$ and $\bar B$. The average value
of asymmetry in the $i^{th}$ bin is then,
\begin{equation}  
\displaystyle A_{i} =
\frac{N_B}{\Gamma_B\,N_i}\int_{q^2_{min}}^{q^2_{max}} dq^2
A(q^2)\frac{d(\Gamma+\bar\Gamma)}{dq^2}\;, 
\end{equation}
where, $N_B$ is the total number of $B$'s, $\Gamma_B$ is the total $B$
width and $q^2_{min}$,$q^2_{max}$ are the minimum and maximum $q^2$
values in the bin; the number of events in the $i^{th}$ bin are,
\begin{equation}  
\displaystyle N_i =
\frac{N_B}{\Gamma_B}\int_{q^2_{min}}^{q^2_{max}} dq^2
\frac{d(\Gamma+\bar\Gamma)}{dq^2}\;. 
\end{equation}
 
The value of the asymmetry, in the $i^{th}$ $q^2$  bin, coming from
$\tilde\kappa$ and $\tilde\lambda$ contributions, is of the form
\begin{equation}
A_i=\frac{a_i \tilde\kappa+b_i \tilde\lambda}{X_i}. 
\end{equation}
In $X_i$, terms quadratic in $\tilde\kappa$ and $\tilde\lambda$ are
ignored, which is a reasonable approximation as long as
$\tilde\kappa$, $\tilde\lambda$ are small.  The observed asymmetry
$A_i^{obs}$ in the $i^{th}$ bin is assumed to be chosen from a
Gaussian distribution with mean equal to the theoretical asymmetry
$A_i^{th}$ and variance $\sigma_i^2$. Then, the method of least
squares gives,
\begin{equation}
 \chi^2=\sum_{i=1}^{nbin}\left(\frac{A_{i}^{obs}-A_{i}^{th}}
 {\sigma_{i}}\right)^2\;.   
\end{equation}
The statistical errors are taken to be $\sigma_{i}=\sqrt{1/N}$. We evaluate the difference,
$\Delta\chi^2=\chi^2_{SM}-\chi^2_{min}$, where $\chi^2_{SM}$ is the
$\chi^2$ corresponding to  the SM values of the parameters i.e.
(0,0) and $\chi_{min}^2$ is evaluated from the values ($\tilde\kappa,
\tilde\lambda$) that minimize $\chi^2$. The linear
dependence on the parameters $\tilde\kappa,\tilde\lambda$ results in
the simple form,
\begin{equation}
 \Delta\chi^2\displaystyle=\displaystyle\sum_{i=1}^{nbin}\displaystyle
\frac{\displaystyle\left(\frac{\displaystyle\tilde\kappa a_i}{X_i}
 \right)^2+\left(\frac{\displaystyle\tilde\lambda b_i}{X_i}
 \right)^2+\frac{\displaystyle 2
 \tilde\lambda \tilde\kappa a_i b_i}{X_i^2}}{\sigma_{i}^2}\;. 
\end{equation}  
Now, $\Delta\chi^2=n^2$ will give the $n$-standard deviation bounds
for $\tilde\kappa$ and $\tilde\lambda$.
        
For 20 bins, for 10,000 $(B \rightarrow K^{*}e^{+}e^{-})$ events, it
is possible to obtain the improved individual bounds, \by
|\tilde\kappa| &<& 0.25 (0.42)\hspace{0.2in} {\rm at}\hspace{0.2in}
68.3(90)\%\; {\rm C.L.} \nn \\ |\tilde\lambda| &<&
0.76(1.25)\hspace{0.2in} {\rm at}\hspace{0.2in} 68.3(90)\%\; {\rm
C.L.}  \ey The possible combined bounds on $\tilde\kappa$ and
$\tilde\lambda$ are shown in Fig. 2.

To conclude, we have studied $C\!P$~ violating anomalous couplings
$\tilde\kappa$ and $\tilde\lambda$ contributing to the process $B
\rightarrow K^{*}\ell^{+}\ell^{-} \to K\pi\ell^+\ell^-$.  These
anomalous couplings can be constrained by constructing an asymmetry,
requiring the addition of $B$ and $\bar B$ events. No strong phases,
nor any flavour/time tagging are required for this
technique. Although, EDM of neutron places strong bounds on the
couplings $\tilde\kappa$ and $\tilde\lambda$, the $C\!P$~ odd
asymmetries studied here, depend on a different combination of these
anomalous couplings and thus provide useful additional
information. $\tilde\lambda$ being the coefficient of a higher
dimensional operator, is expected to be better constrained by high
energy collider events than from rare B decays. The recent results of
D0 provide good bounds on $\tilde\lambda$, but for $\tilde\kappa$ the
approach discussed here can provide a much tighter constraint.

We would like to thank G.Date and G.Rajasekaran for discussions.

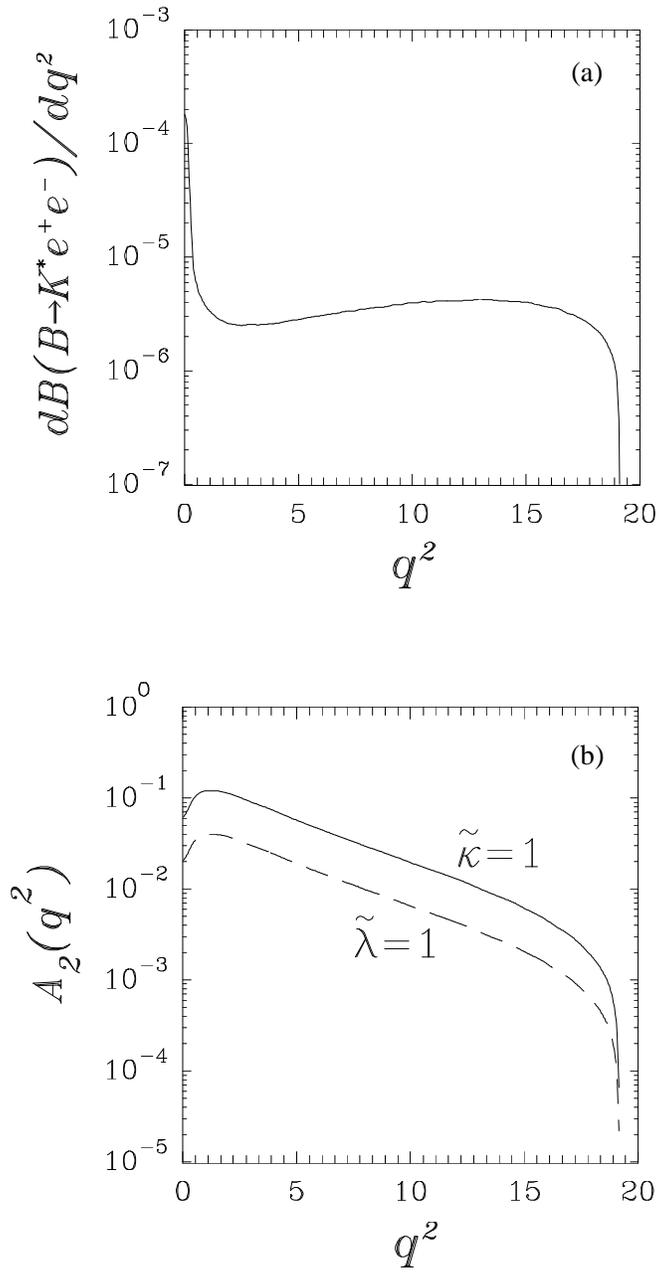
\begin{figure}
\caption{
(a)~The differential branching fraction as a function of $q^2$. (b)~The  asymmetry $A_2(q^2)$ as a function of $q^2$. The solid line
is for $\tilde\kappa=1(\tilde\lambda=0)$ and the dashed for
$\tilde\lambda=1(\tilde\kappa=0)$. 
}\label {Fig. 1}
\end{figure}

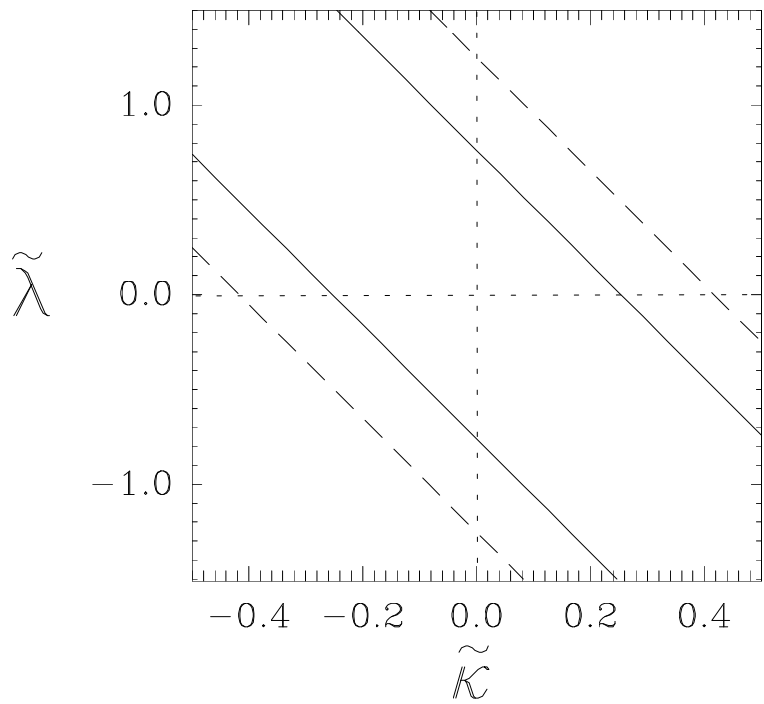
\begin{figure}
\caption{
Limits on the $C\!P$~-violating anomalous coupling parameters $\tilde\kappa$
and $\tilde\lambda$. The region inside the solid (dashed) lines
represents the 68.3(90)\% C.L. limits.
}\label {Fig. 2}
\end{figure}

\begin{table}
\caption{ Relations between the form factors used in this paper,
a quark model (QM) that reproduces heavy quark limit . $\displaystyle
W_\mu=(K_\mu-\zeta\,k_\mu)$, $\sigma^2=96  
\pi^2/(m_{K^*}^2 {\lambda_K}^3)$, $\zeta=\displaystyle\frac{k\cdot
K}{m_{K^*}^2}$, $\lambda_K$ and $~\Delta$ are defined in the text.
}
\label{tab1}
\begin{tabular}{rcl}
{~~~~~~~~~~~~~~~~~~~~~~~~~~~~~~~~}& QM
\cite{Wyler}&{~~~~~~~~~~~~~~~~~~~~}\\
\hline
${\cal A}$ & $-2\,g\,\sigma$    \\[0ex]
${\cal B}$ & $a_0\,\Delta\,\sigma$  \\[2ex] 
${\cal C}$ & $\displaystyle 2\,a_{+}\,W\cdot q\,\sigma-\zeta\,{\cal B}$   \\ [2ex] 
${\cal D}$ & $\displaystyle -2\,\frac{V}{M_B+m_{K^*}}\,\sigma $  \\[2ex]
${\cal E}$ & $\displaystyle \frac{A_0}{M_B+m_{K^*}}\, \Delta\, \sigma $ 
 \\[1ex]
${\cal F}$ & $\displaystyle 2\,\frac{A_{+}}{M_B+m_{K^*}}\,W\cdot 
q\,\sigma-\zeta
   \,{\cal E} $\\ [2ex]
\end{tabular}      
\end{table}

\end{document}